\documentstyle[fleqn]{tp}

\input psfig

\def\etal{{\it et\thinspace al.\ }}

\def\ga{\mathrel{\mathchoice {\vcenter{\offinterlineskip\halign{\hfil
$\displaystyle##$\hfil\cr>\cr\sim\cr}}}
{\vcenter{\offinterlineskip\halign{\hfil$\textstyle##$\hfil\cr>\cr\sim\cr}}}
{\vcenter{\offinterlineskip\halign{\hfil$\scriptstyle##$\hfil\cr>\cr\sim\cr}}}
{\vcenter{\offinterlineskip\halign{\hfil$\scriptscriptstyle##$\hfil
\cr>\cr\sim\cr}}}}}
\def\la{\mathrel{\mathchoice {\vcenter{\offinterlineskip\halign{\hfil
$\displaystyle##$\hfil\cr<\cr\sim\cr}}}
{\vcenter{\offinterlineskip\halign{\hfil$\textstyle##$\hfil\cr<\cr\sim\cr}}}
{\vcenter{\offinterlineskip\halign{\hfil$\scriptstyle##$\hfil\cr<\cr\sim\cr}}}
{\vcenter{\offinterlineskip\halign{\hfil$\scriptscriptstyle##$\hfil
\cr<\cr\sim\cr}}}}}

\newcommand{\tF}{\tilde{F}}

\newcommand{\tA}{\tilde{A}}

\newcommand{\tS}{\tilde{S}}

\renewcommand{\bar}{\overline }

\newcommand{\xiav}{\bar{\xi}}

\begin{document}

\twocolumn[
\title{The Cosmic Distribution of Clustering}
\author{S. Colombi$^1$, I. Szapudi$^2$ and the VIRGO
consortium$^3$\\
{\it $^1$Institut d'Astrophysique de Paris, CNRS,}\\
{\it 98bis bd Arago,F-75014 Paris, France}\\
{\it $^2$University of Durham, Department of Physics,}\\ 
{\it South Road, Durham, DH1 3LE, UK}\\
{\it $^3$See, e.g., http://star-www.dur.ac.uk/$^{\sim}$frazerp/virgo/people.html}}
\vspace*{16pt}   

ABSTRACT.\
For a given statistic, $A$, the {\em cosmic distribution function}, 
$\Upsilon(\tA)$, is the probability of measuring a value $\tA$ 
in a finite galaxy catalog. For statistics related to 
count-in-cells, such as factorial moments, $F_k$, 
the average correlation function, $\xiav$, and cumulants,
$S_N$, the functions $\Upsilon(\tF_k)$,
$\Upsilon(\tilde{\xiav})$, and
$\Upsilon(\tS_N)$ were measured in a large $\tau$CDM simulation.
This $N$-body experiment simulates almost the full ``Hubble Volume''
of the universe, thus, for the first time, it
allowed for an accurate analysis of the cosmic distribution function, and, 
in particular, of its variance $(\Delta A)^2$,
the {\em cosmic error}. The resulting detailed knowledge about the 
shape of $\Upsilon$ is crucial for likelihood analyses.
The measured  cosmic error agrees remarkably well with the theoretical
predictions of Szapudi \& Colombi (1996) and Szapudi, Bernardeau \&
Colombi (1998) in the weakly non-linear regime, while the predictions
are slightly above the measurements in the highly nonlinear regime.
When the relative cosmic error is small,  
$(\Delta A/A)^2 \ll 1$, function $\Upsilon$ is nearly
Gaussian. When $(\Delta A/A)^2$ approaches unity or is larger,
function $\Upsilon(\tA)$ is increasingly skewed and well
approximated by a lognormal distribution for $A=F_k$, or $A=\xiav$.
The measured cumulants follow accurately the perturbation theory predictions
in the weakly nonlinear regime. Extended perturbation theory is an
excellent approximation for all the available dynamic range.
\endabstract]

\markboth{S. Colombi, I. Szapudi \etal}{Cosmic Distribution}

\small

\section{Introduction}
To confront theory with observations, an accurate understanding of errors is 
essential.
The state of the art maximum likelihood approach 
requires full information on the distribution function of measurements;
the scatter alone is insufficient if the underlying error distribution
is non-Gaussian. As shown later, this is often the case in
large scale structure studies. What follows, focuses
on statistics related to count-in-cells (CIC),
in particular, factorial moments, $A=F_k$ (e.g., Szapudi \& Szalay,
1993), the averaged two point correlation function, $A=\xiav$, 
and higher order cumulants, $A=S_N$ (e.g., Balian \& Schaeffer, 1989).
We are concerned with the question: 
what is the probability of measuring
a value $A=\tA$ in a finite galaxy catalog,  ${\cal E}_i$, with
volume $V$? The CIC indicators from a particular realization
are influenced by various statistical effects (e.g.,
Szapudi \& Colombi 1996, hereafter SC), which can be approximately
separated:
\begin{enumerate}
\item Finite volume effects are due to fluctuations of
the density field on scales larger than the catalog.
\item Edge effects are caused by the uneven statistical weight
given to objects near the survey boundary.
\item Discreteness effects are related to the finite number
of galaxies in the catalog, which sample the underlying continuous
field. 
\end{enumerate}
In a large number of realizations, $1 \leq i
\leq C_{\cal E}$,  the distribution of measurements, 
the {\em cosmic distribution function} $\Upsilon(\tA)$ could
be estimated. In most cases, this is unachievable in practice,
but  $\Upsilon(\tA)$ could still be evaluated theoretically,
or via simulations. An important special case is the
Gaussian distribution, which is fully characterized by
its average, $\langle \tA \rangle \equiv
A$, and its variance,  $(\Delta A/A)^2$, the {\em
cosmic error}. 

This contribution concentrates on the properties $\Upsilon$, 
representing a fraction of the
results from a more extended article by  Colombi, Szapudi, \etal 1998
based on the latest $N$-body experiment  of the VIRGO consortium.
The next section compares the measured
cumulants,  $S_N$, $1 \leq N \leq 10$, with perturbation theory (PT, e.g., 
Juszkiewicz \etal 1993; Bernardeau, 1994) and extended
perturbation theory predictions (EPT, e.g., Colombi \etal 1996). 
Section $\S 3$ collates the numerical cosmic
error with the theoretical predictions
of SC, and Szapudi, Bernardeau \& Colombi (1998, hereafter SBC). Finally,
Section $\S 4$ exposes the shape of cosmic distribution function 
$\Upsilon$.

\section{The Underlying Statistics}
The algorithm of Szapudi \etal (1998) was employed to 
extract CICs from a
$\tau$CDM simulation with one billion particles
in a cubic box of $2000\,h^{-1}$ Mpc (see the contribution of
A. Evrard in the same volume). Note that most of the considerations
of this work are quite insensitive to the particular
cosmological model, thus similar results are expected for all
currently fashionable CDM variants.
The virtual (periodic) universe was divided 
into $C_{\cal E}=16^3$ adjacent cubic subsamples, each of them
representing a possible realization of our visible, local
universe. \footnote{The box size was conveniently chosen for
the compressed output format provided by Adrian Jenkins 
to speed up processing.}
$512^3$ cubical sampling cells of size $\ell$ 
were placed in the full simulation, and in each subsample ${\cal E}_i$,
covering the scale range of 
$0.24 h^{-1} \leq \ell \leq 250 h^{-1}$ Mpc. 
The combined subsamples probed 
the tail of the CIC probability distribution function, $P_N$, extremely
well, down to $P_N \simeq 1.8.10^{-12}$. The 
measurement in the whole sample is somewhat less accurate, but still
state of the art, $P_N \ga 7.5.10^{-9}$. 

\begin{figure*}[htb]
\centering\mbox{\psfig{figure=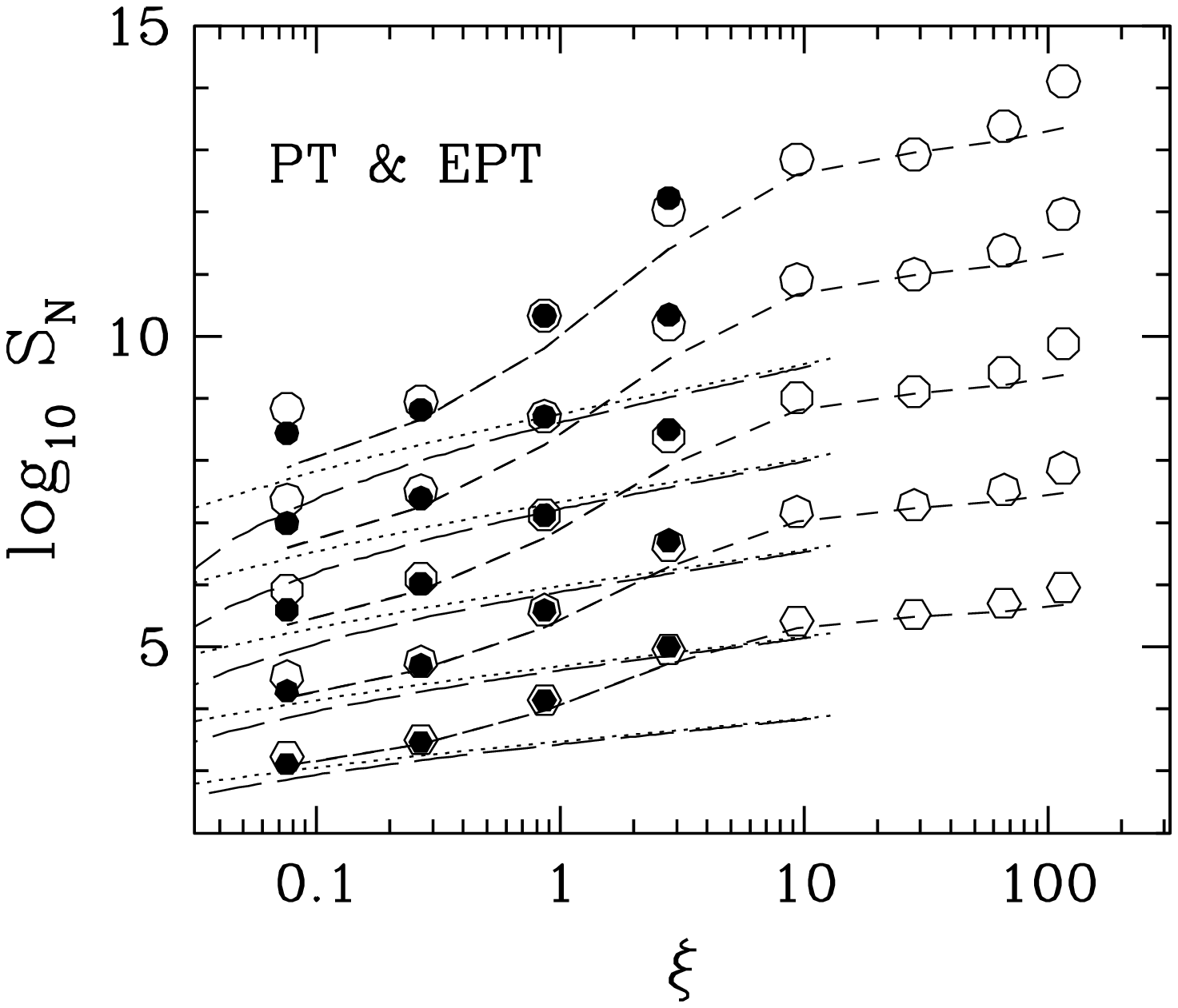,height=5cm}}
\centering\mbox{\psfig{figure=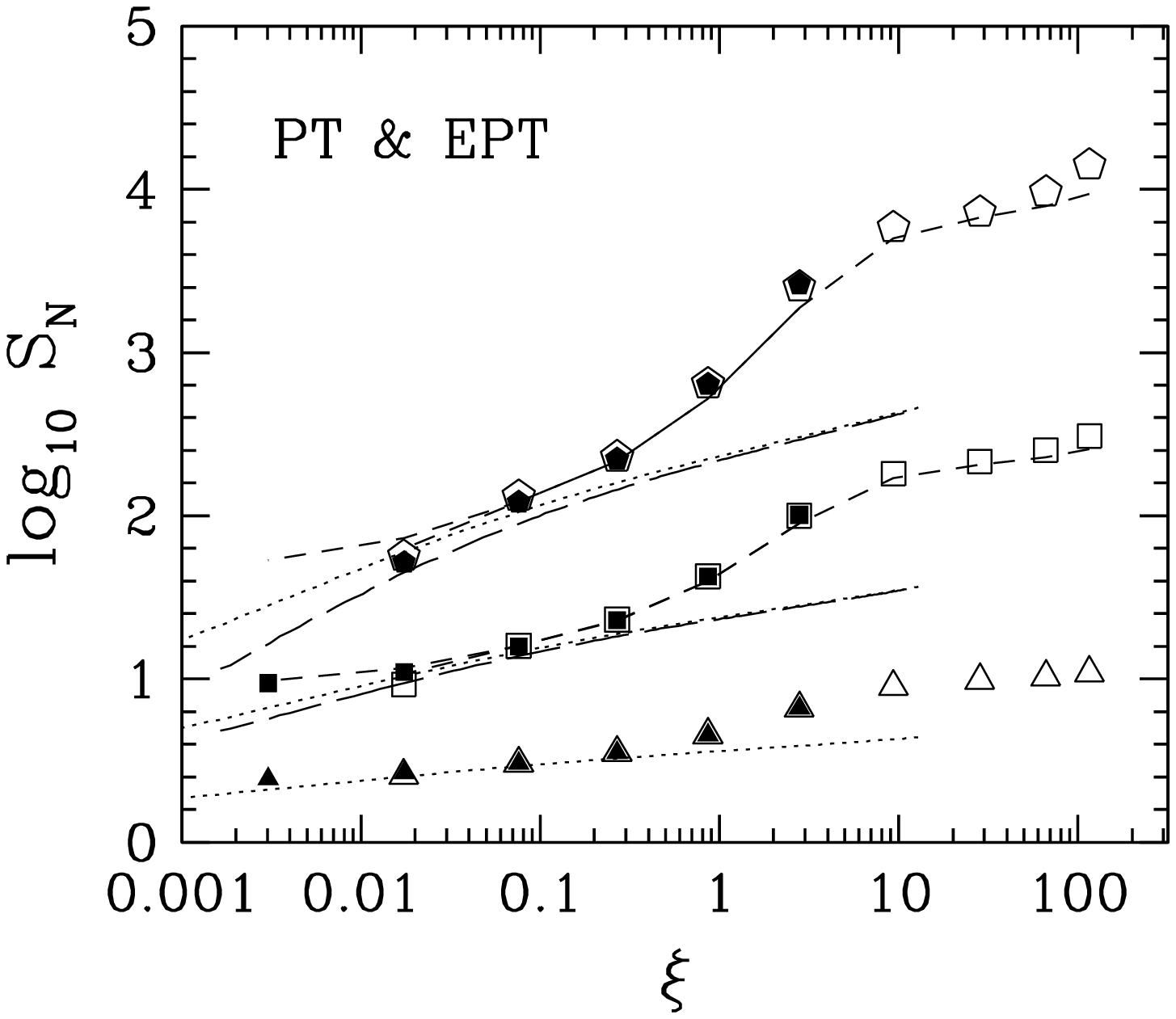,height=5cm}}
\caption[]{The cumulants, $S_N$, measured in the whole $\tau$CDM simulation
are shown as functions of the variance
$\xiav$.  The dots, long dashes
and short dashes correspond to PT with only terms
corresponding to the first order logarithmic derivative of $\xiav$, PT
with only terms up to the second order logarithmic derivative of $\xiav$,
and EPT (see, e.g., Bernardeau 1994 and Colombi \etal 1996), respectively. 
The lower, and upper panels correspond to $3 \leq N \leq 5$, and 
to $6 \leq N \leq 10$, respectively. The $S_N$'s are increasing with
$N$. Finally, the filled, and open symbols correspond to the 
results from the full simulation, and from the combination of 
all the subsamples ${\cal E}_i$,  respectively.}
\label{fig:figure1}
\end{figure*}

Figure~\ref{fig:figure1} shows the cumulants as functions
of the variance, representing the most accurate measurement 
in the widest dynamic range to date.
In agreement with previous studies (see, e.g.,
Gazta\~naga \& Baugh, 1995, Szapudi \etal 1998), the results
match  PT extremely well in the weakly nonlinear regime 
where $\xiav \la 1$, and
EPT is an excellent approximation in all the available dynamic range. 
The 1-loop calculations (not displayed on figure~\ref{fig:figure1}) 
based on spherical model by Fosalba \& Gazta\~naga (1998) show good agreement
 with the numerical results up to $\xiav =1$. 

\section{The Cosmic Error}

\begin{figure*}
\centerline{\mbox{\psfig{figure=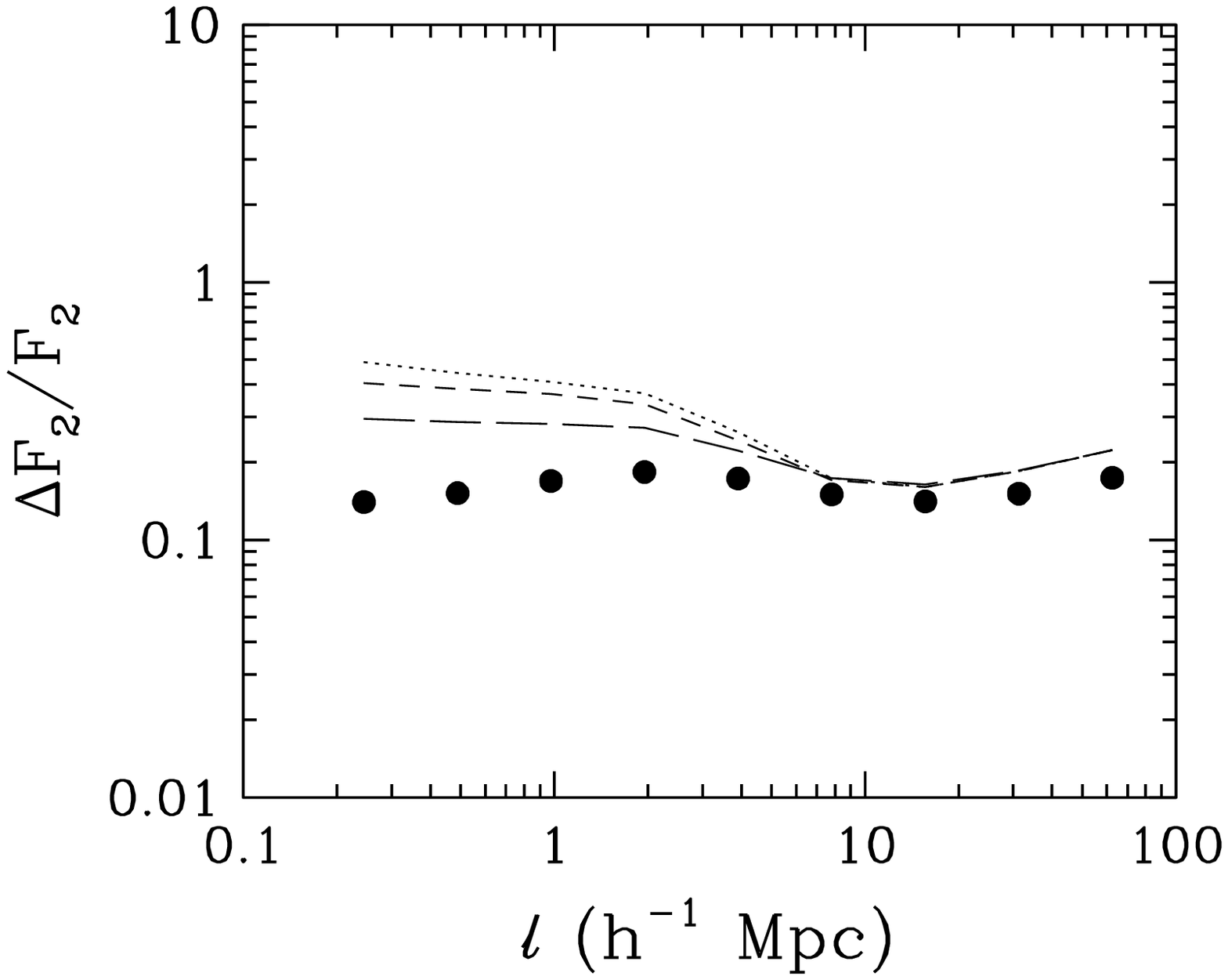,width=4.4cm}\psfig{figure=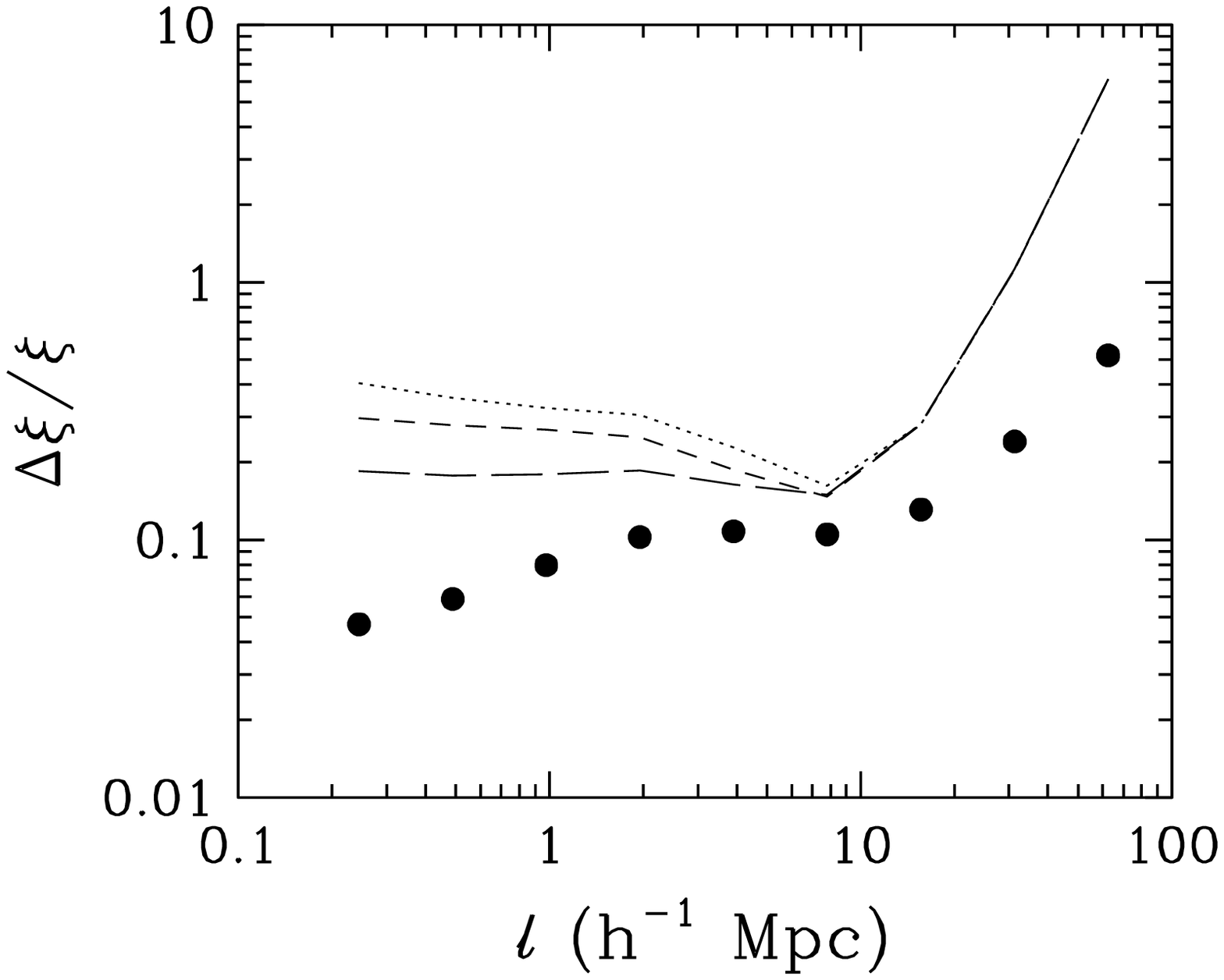,bbllx=57pt,bblly=111pt,bburx=533pt,bbury=496pt,width=4.4cm}}}
\centerline{\mbox{\psfig{figure=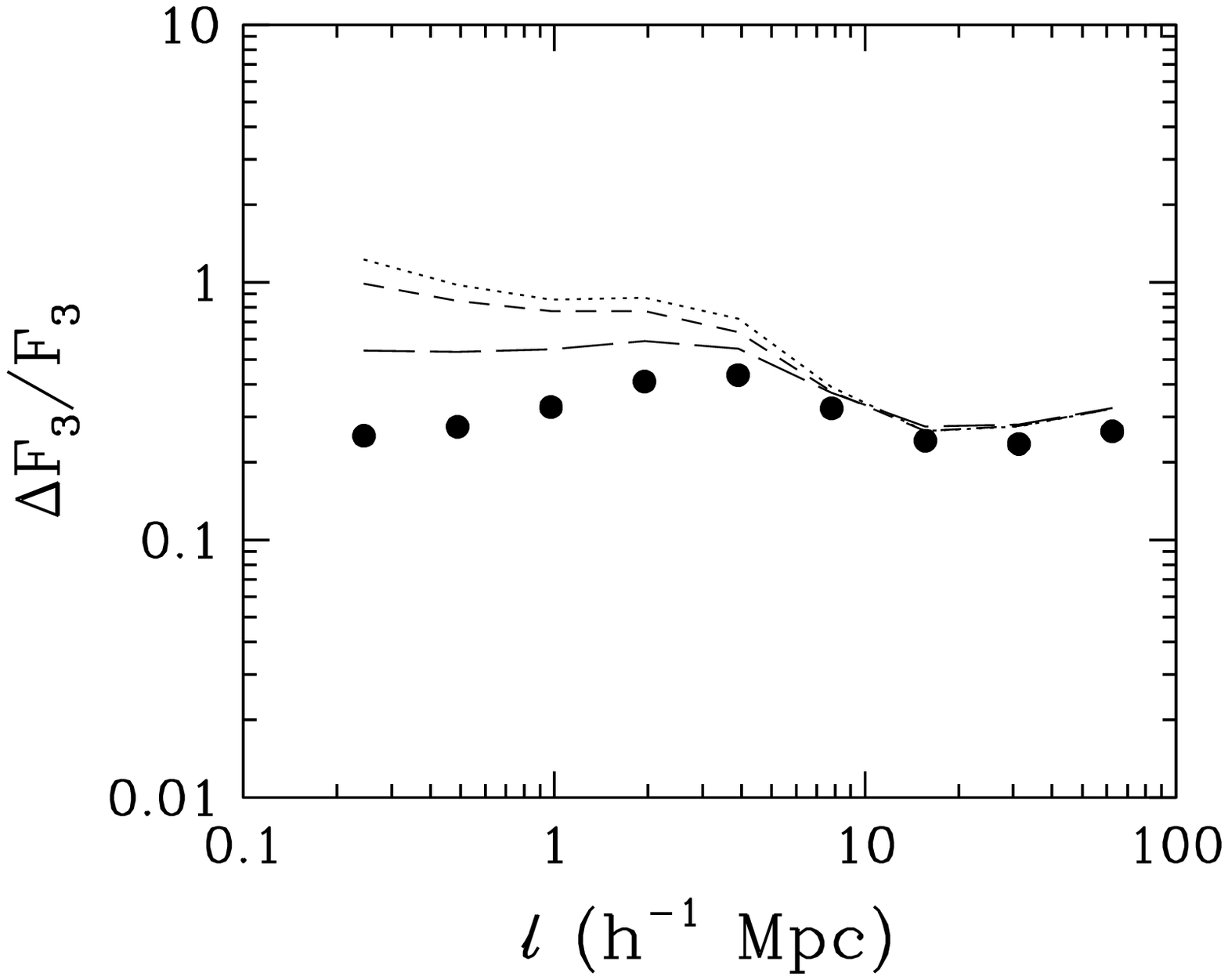,width=4.4cm}\psfig{figure=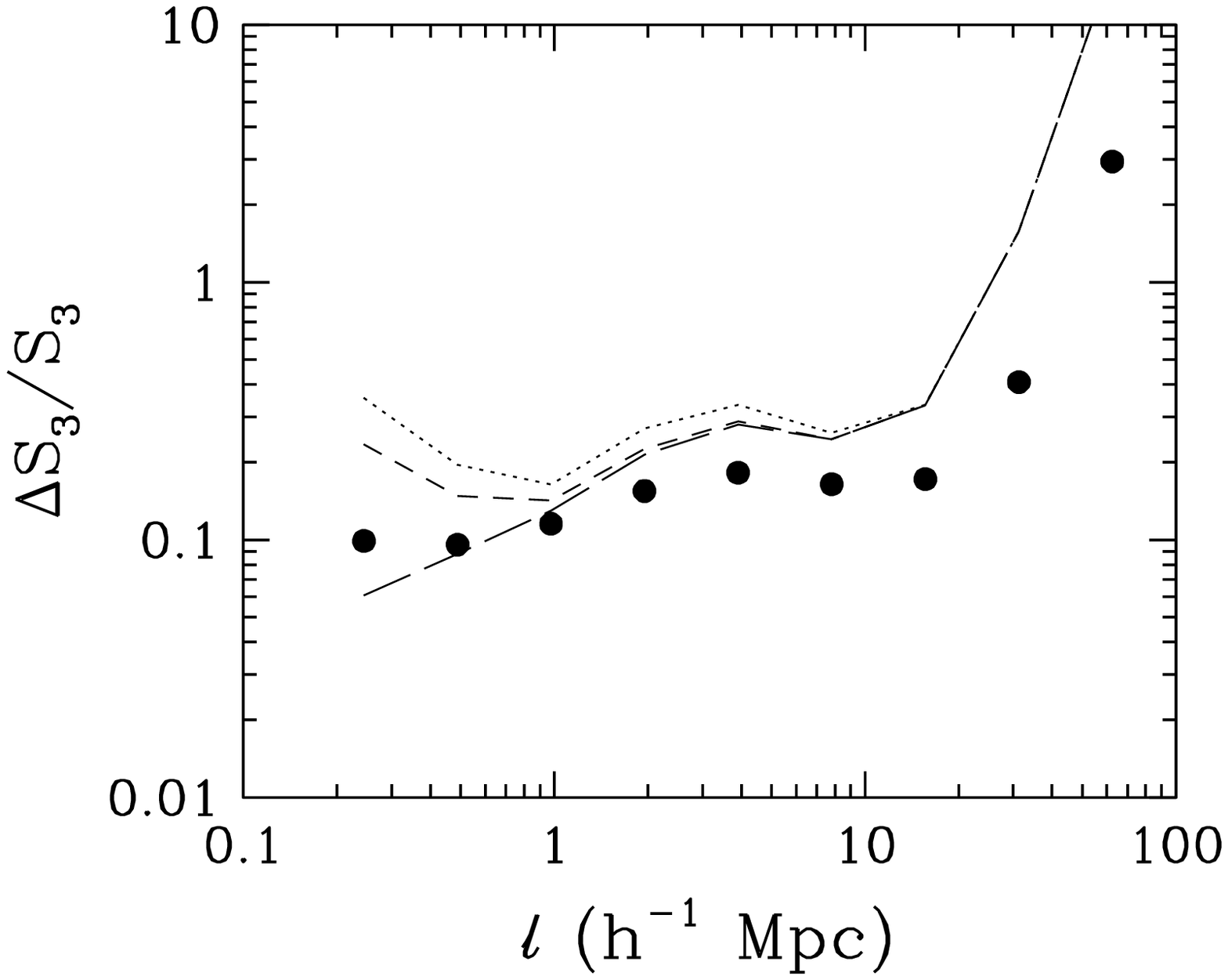,bbllx=57pt,bblly=111pt,bburx=533pt,bbury=496pt,width=4.4cm}}}
\centerline{\mbox{\psfig{figure=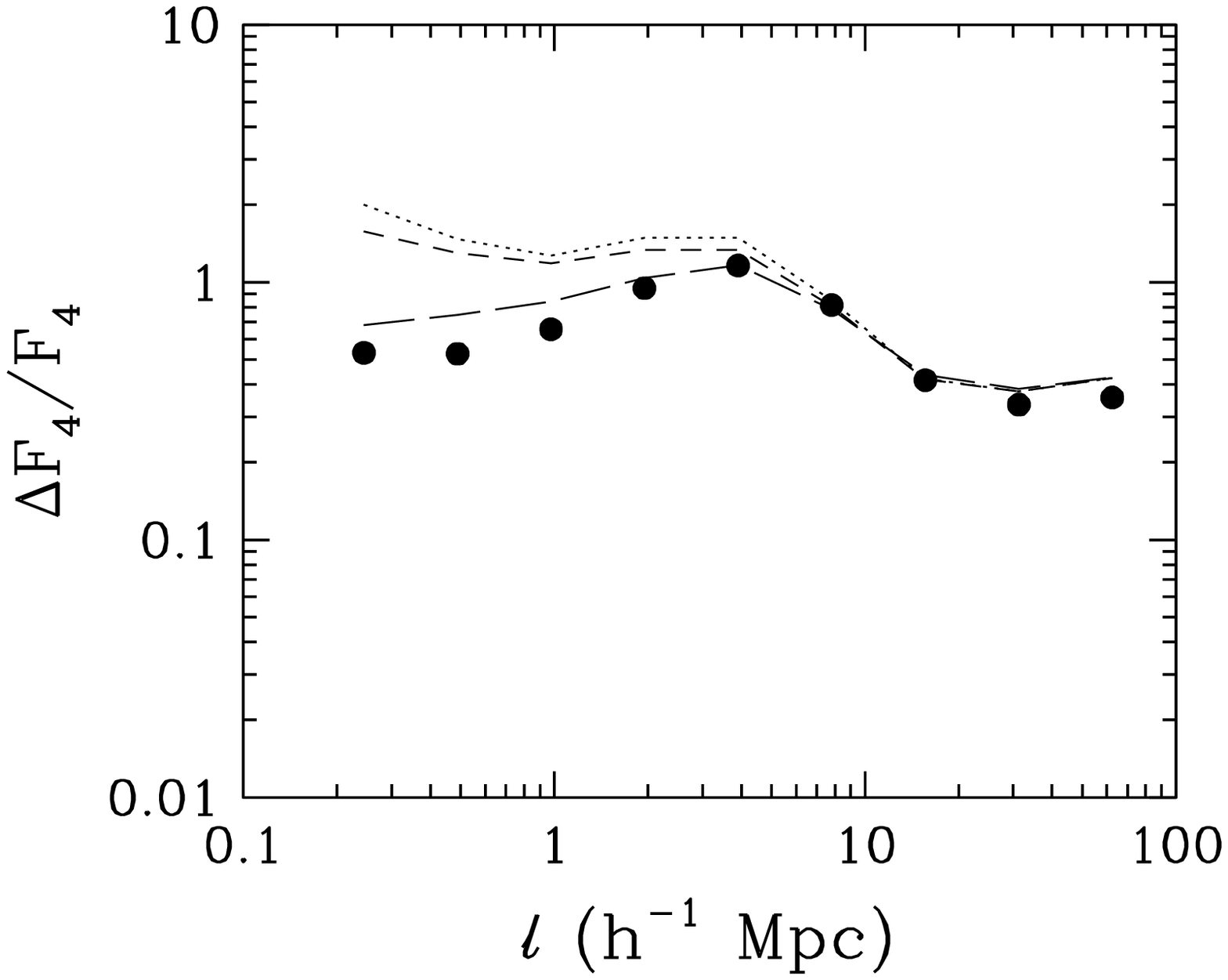,width=4.4cm}\psfig{figure=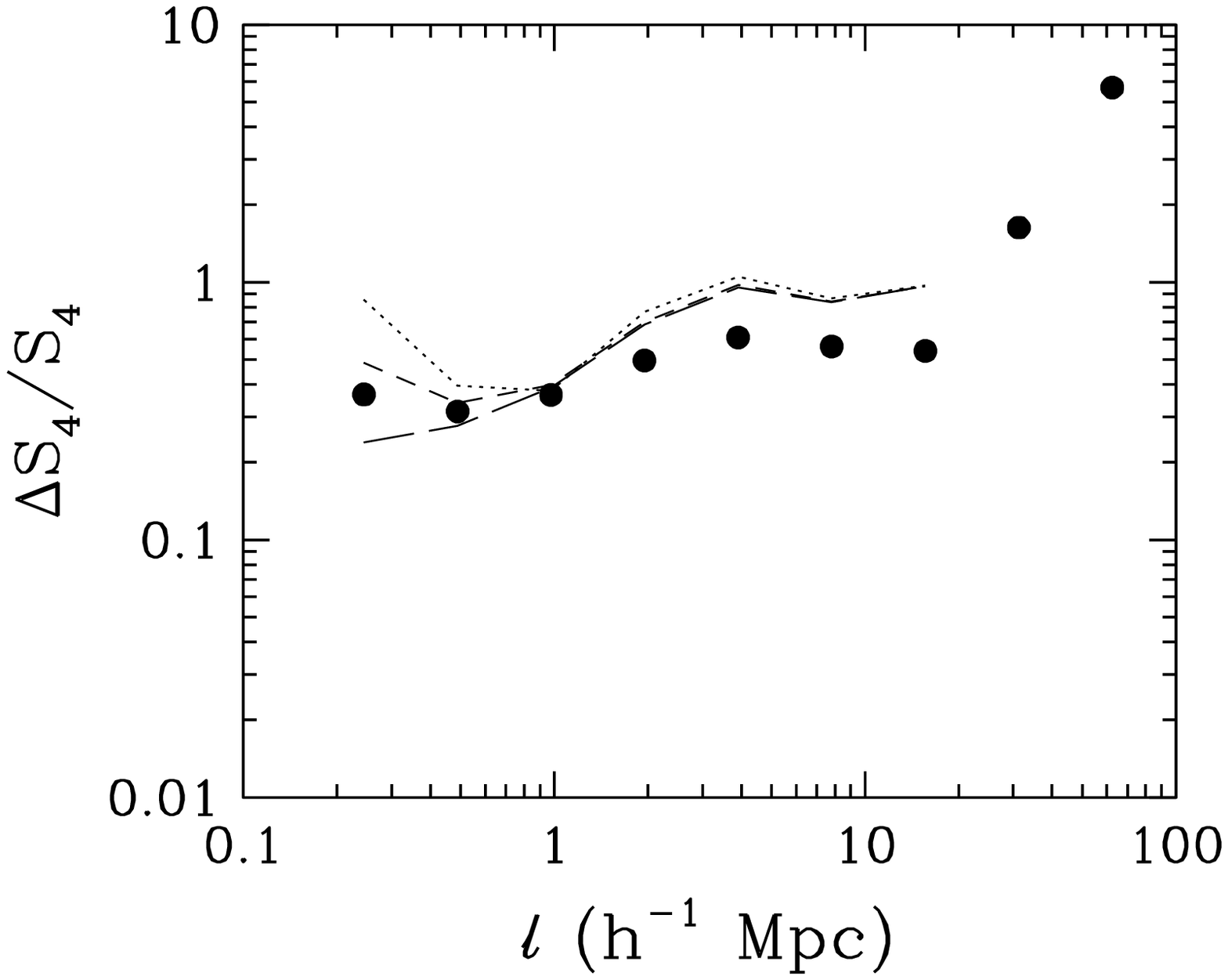,bbllx=57pt,bblly=111pt,bburx=533pt,bbury=496pt,width=4.4cm}}}
\caption[]{The relative cosmic error as a function of scale for the
factorial moments (left panels), $\xiav$, and the cumulants (right
panels) are shown. The symbols represent the cosmic scatter obtained from the
measurements in ${\cal E}_i$, $i=1,\ldots,16^3$. The dots, dashes and long
dashes display the theoretical predictions, based
on the hierarchical models of Szapudi \& Szalay (1993), Bernardeau \&
Schaeffer (1992), and EPT (SBC), respectively. Note that when
the cosmic error approaches unity or, is larger, the theoretical
calculation is not expected to be valid in the right panels, as
it relies on a Taylor expansion in terms of the relative error.}
\label{fig:figure2}
\end{figure*}

Figure~\ref{fig:figure2} shows the measured cosmic error together with the
predictions of SC and SBC. The various theoretical models
agree extremely well with the measurements
in the weakly nonlinear regime (at least for the $F_k$'s) and
tend to be slightly higher than the numerical estimates 
in the nonlinear regime. EPT yields the closest match to the data. 
Taken at face value, these results indicate that
the hierarchical assumption for the joint moments used 
by SC and SBC is inaccurate on the smallest scales, and should
be corrected for the error calculations, where bivariate distributions
play a crucial role. 

It is interesting although not surprising to note that factorial moments
and cumulants of the same order behave differently in terms
of errors. For example, on large scales, where the cosmic error is
dominated by edge effects (SC), the cumulants, $\xiav$, and $S_3$ have
larger scatter than the full moments, $F_2$, and $F_3$;
the reverse is true on small scales. 
The situation is exactly analogous to the ''integral constraint'' problem.

\section{The Cosmic Distribution Function}

\begin{figure*}
\centerline{\mbox{
\psfig{figure=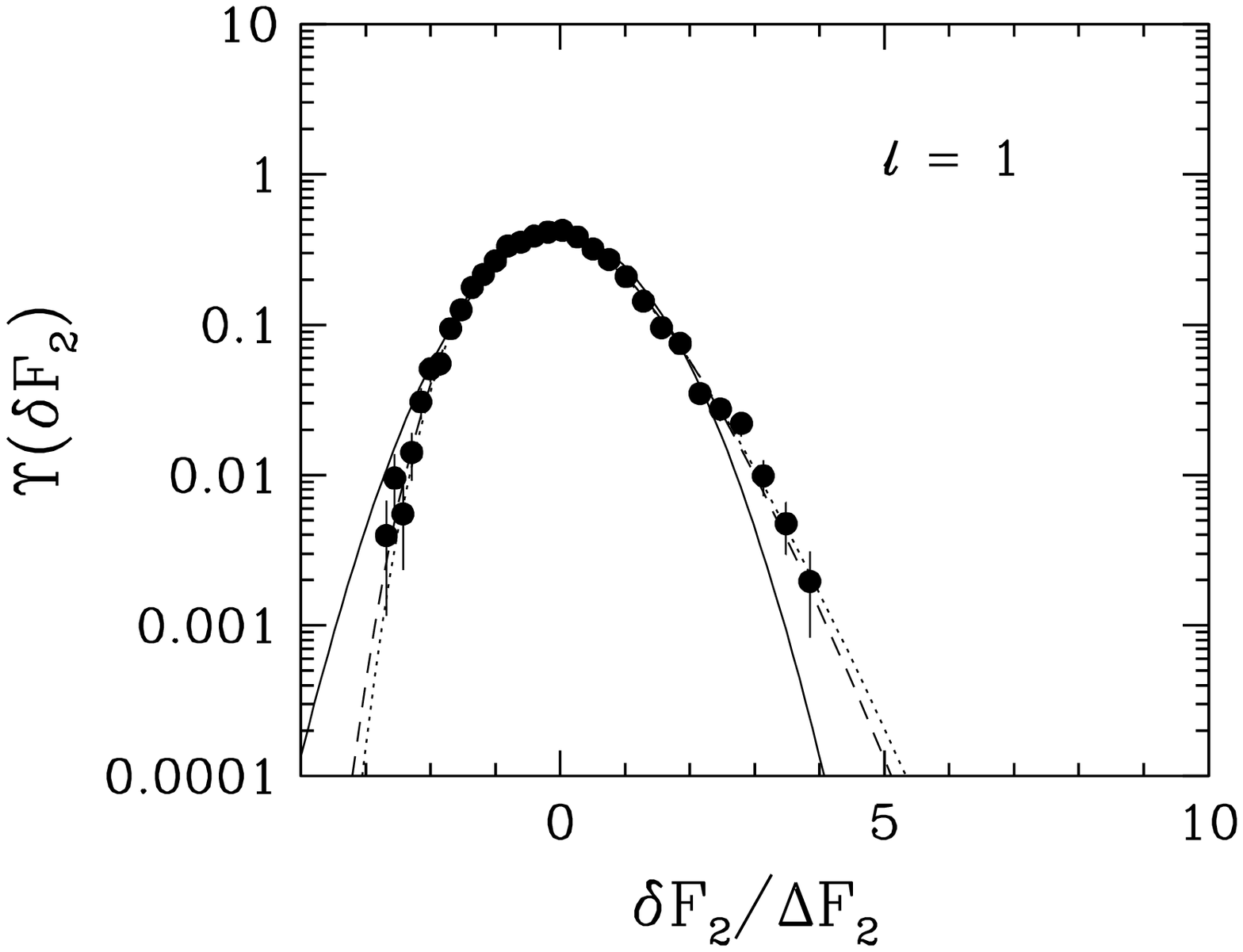,bbllx=35pt,bblly=114pt,bburx=527pt,bbury=492pt,width=4cm}\psfig{figure=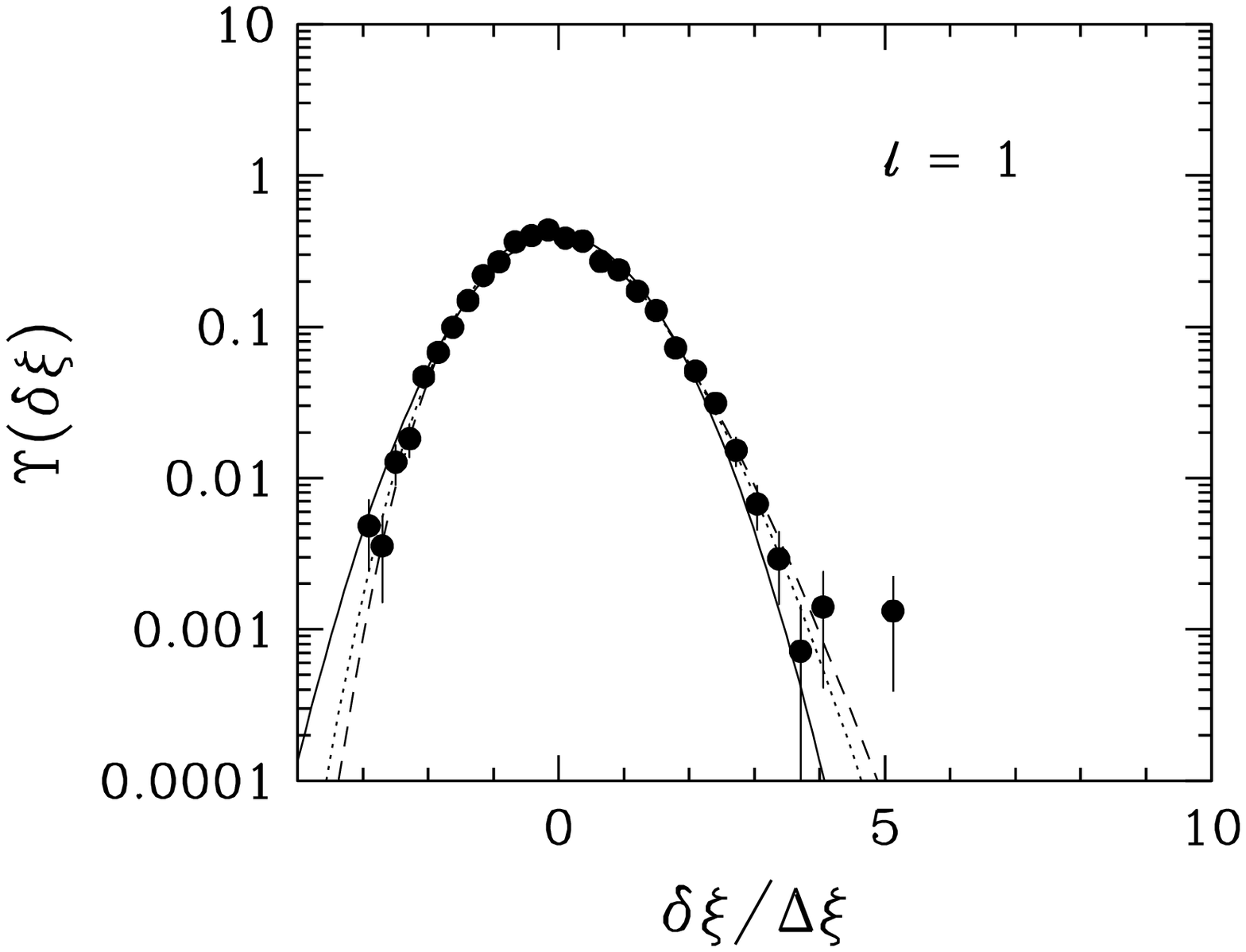,bbllx=35pt,bblly=114pt,bburx=527pt,bbury=492pt,width=4cm}}}
\centerline{\mbox{
\psfig{figure=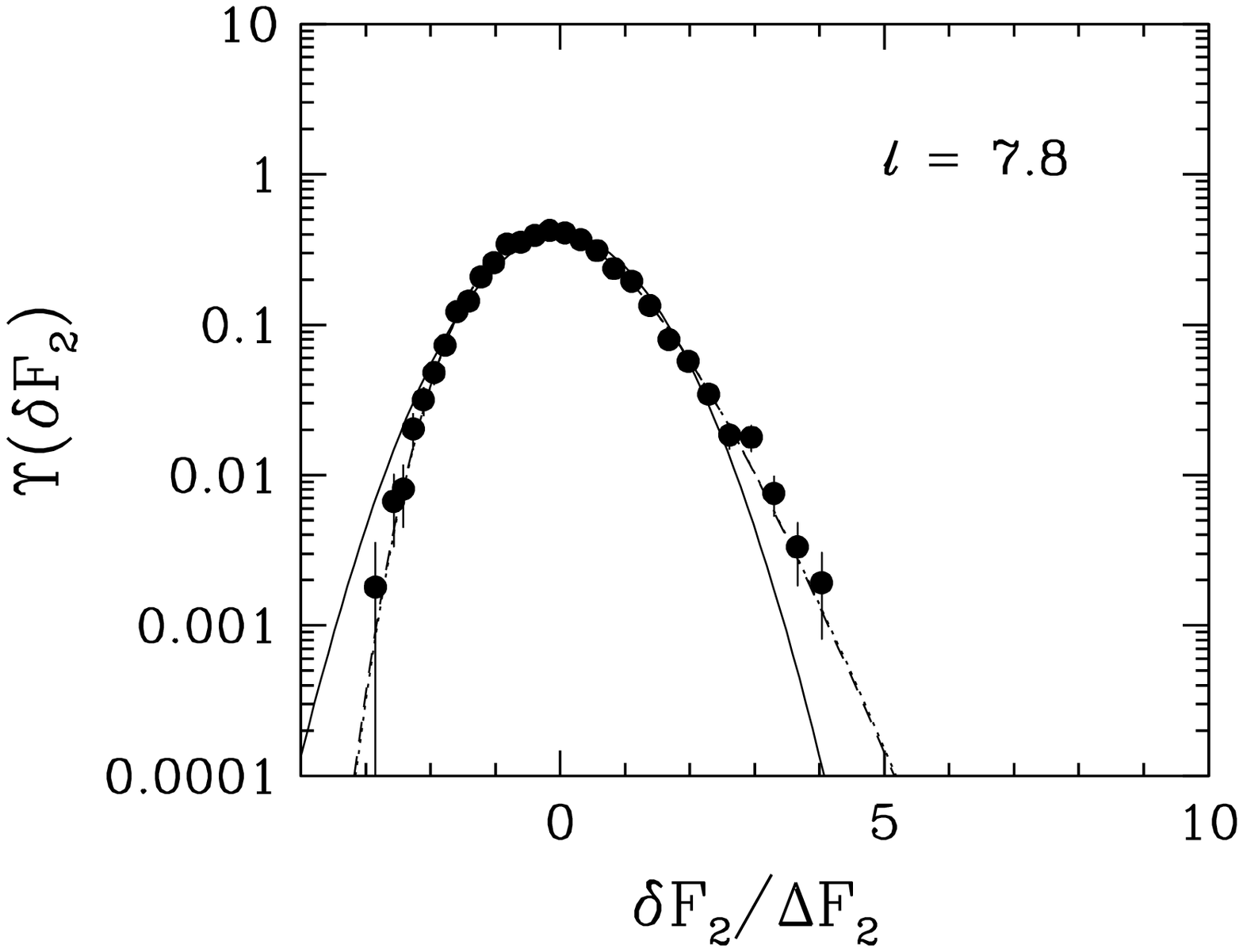,bbllx=35pt,bblly=114pt,bburx=527pt,bbury=492pt,width=4cm}\psfig{figure=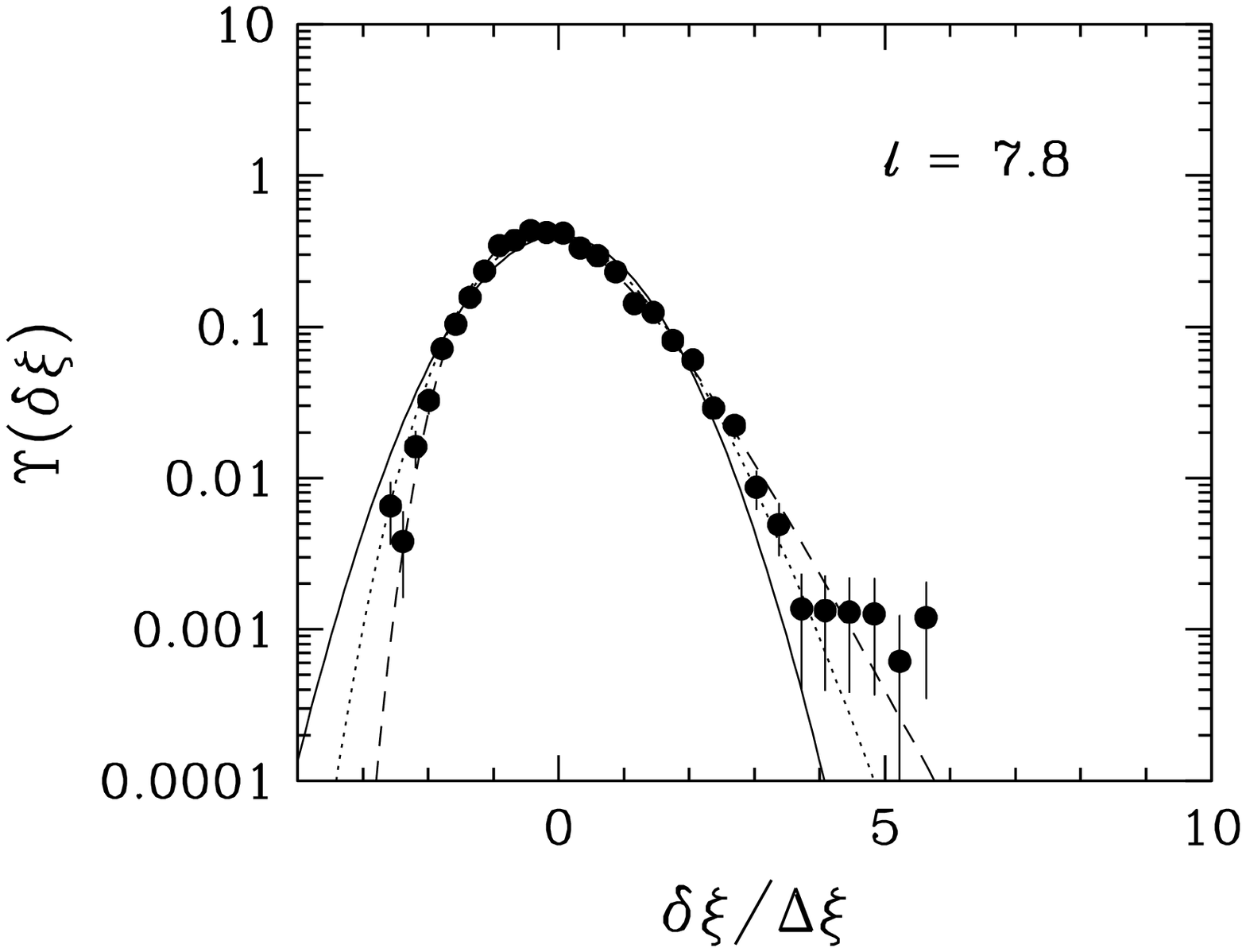,bbllx=35pt,bblly=114pt,bburx=527pt,bbury=492pt,width=4cm}}}
\centerline{\mbox{\psfig{figure=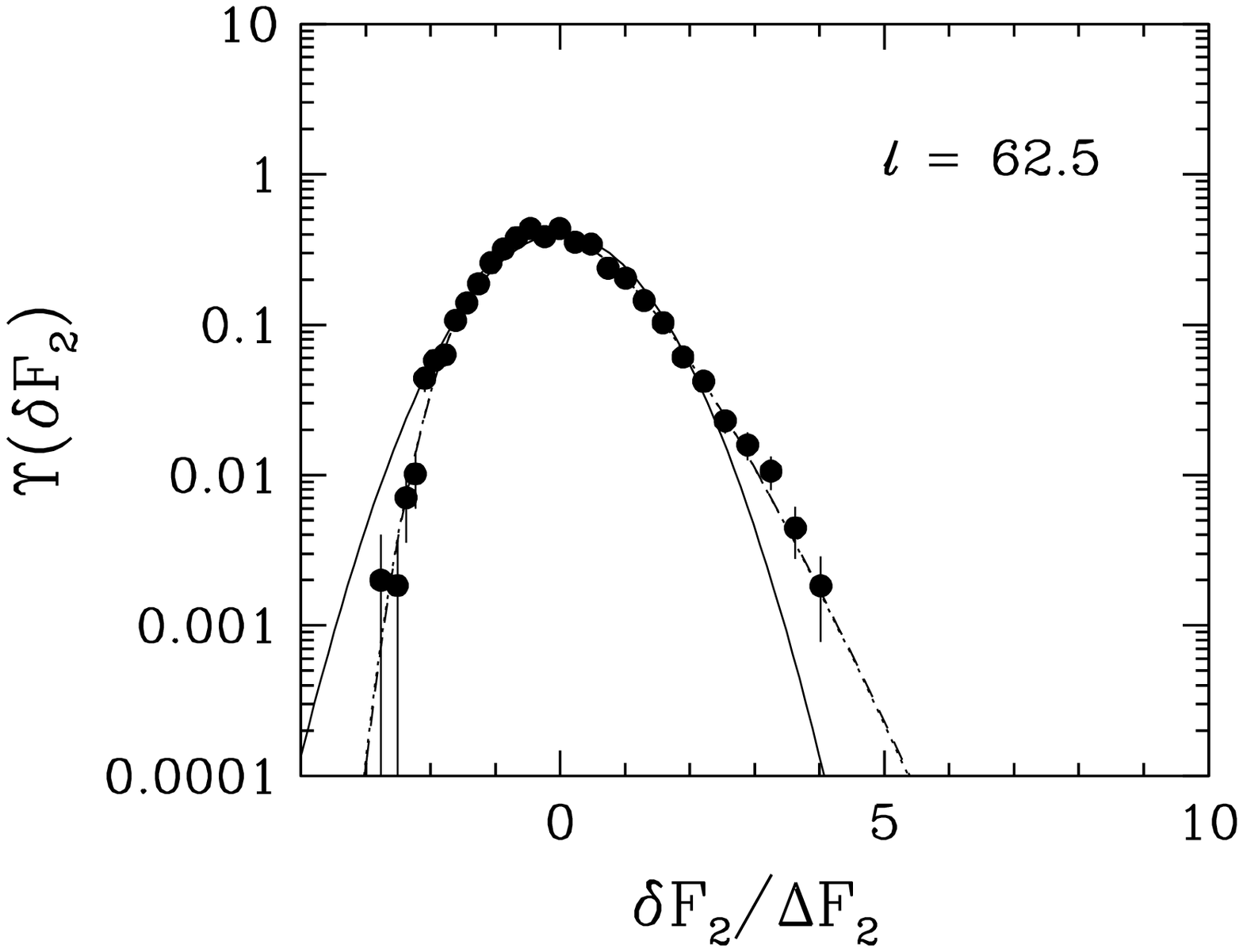,bbllx=35pt,bblly=114pt,bburx=527pt,bbury=492pt,width=4cm}\psfig{figure=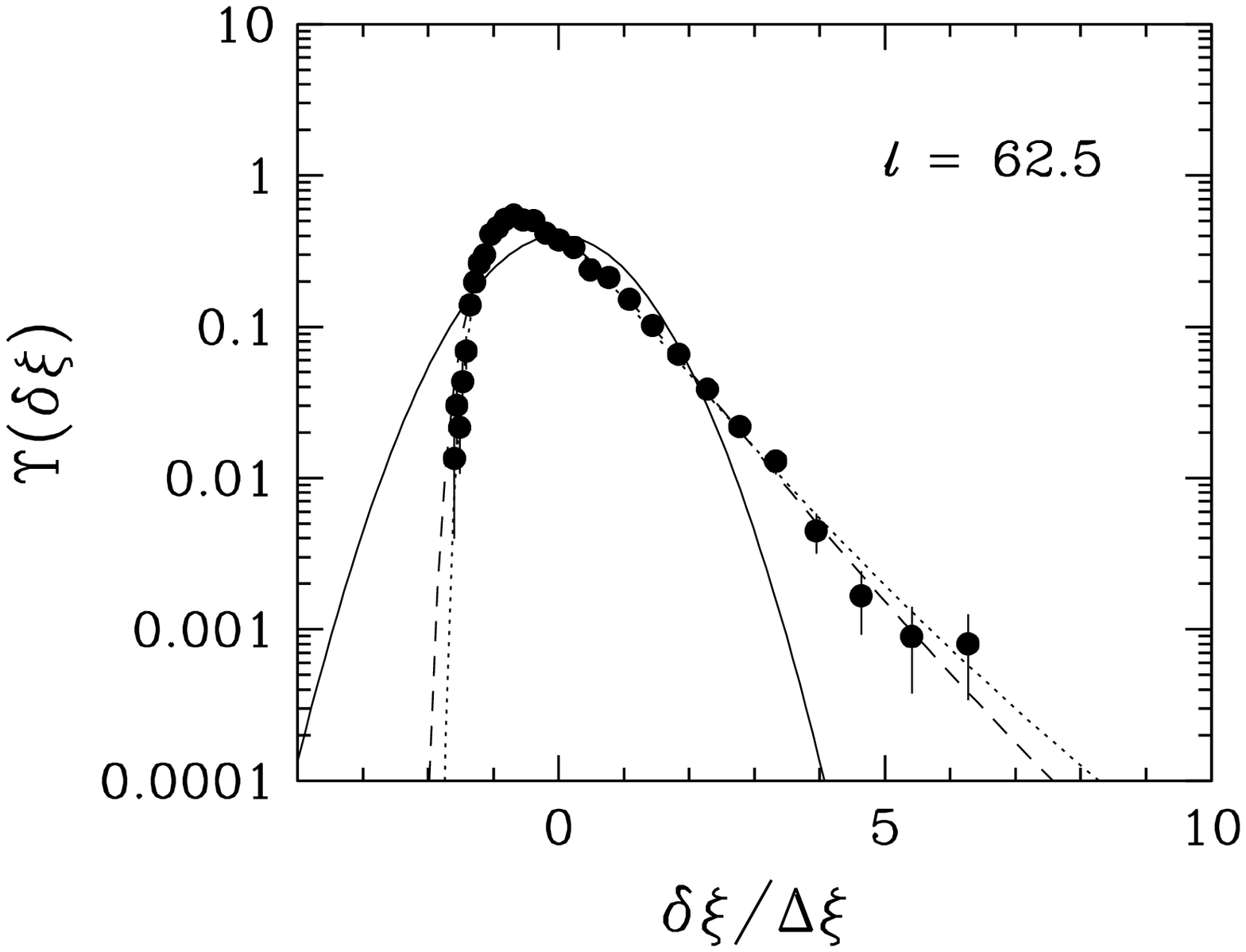,bbllx=35pt,bblly=114pt,bburx=527pt,bbury=492pt,width=4cm}}}
\centerline{\mbox{
\psfig{figure=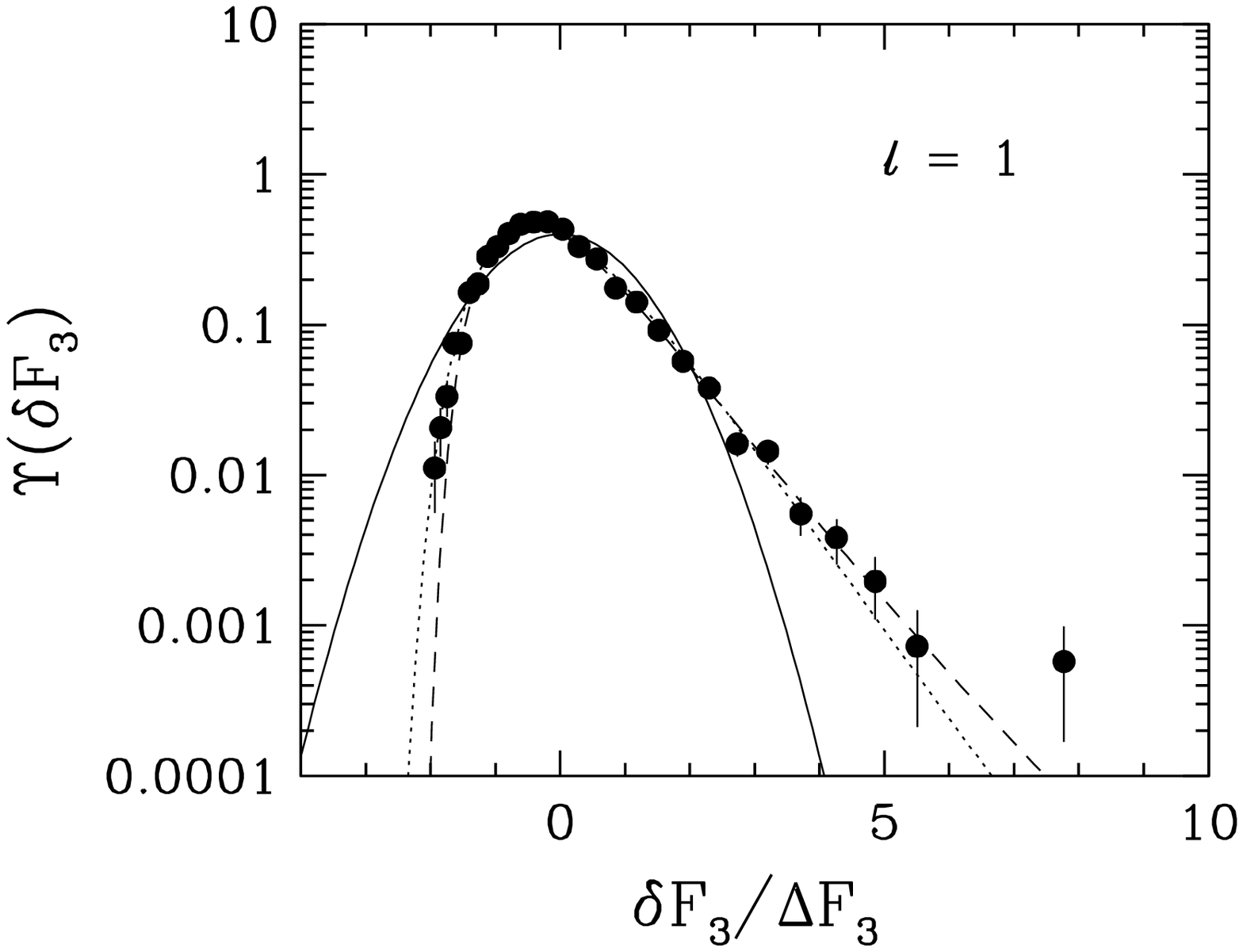,bbllx=35pt,bblly=114pt,bburx=527pt,bbury=492pt,width=4cm}\psfig{figure=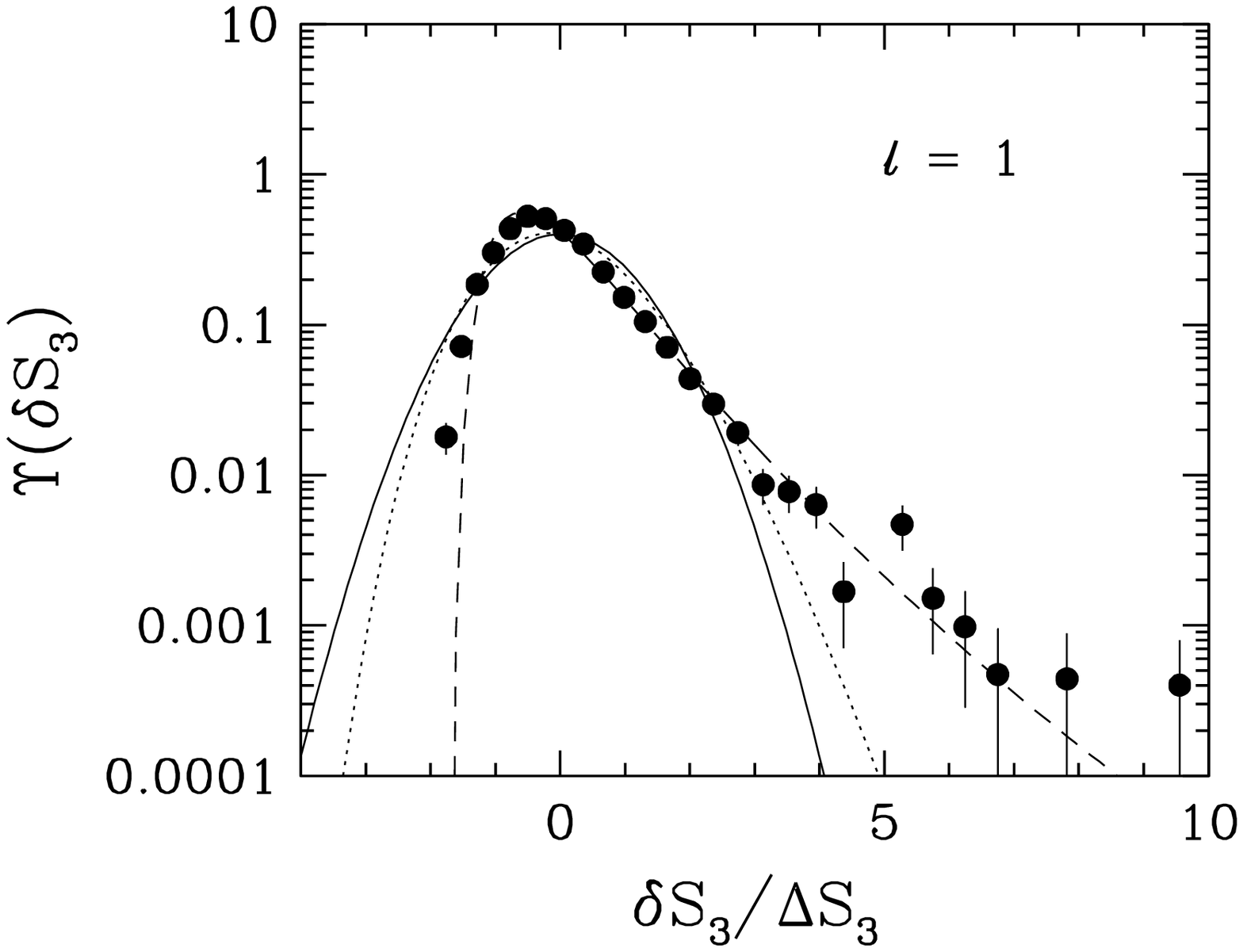,bbllx=35pt,bblly=114pt,bburx=527pt,bbury=492pt,width=4cm}}}
\centerline{\mbox{
\psfig{figure=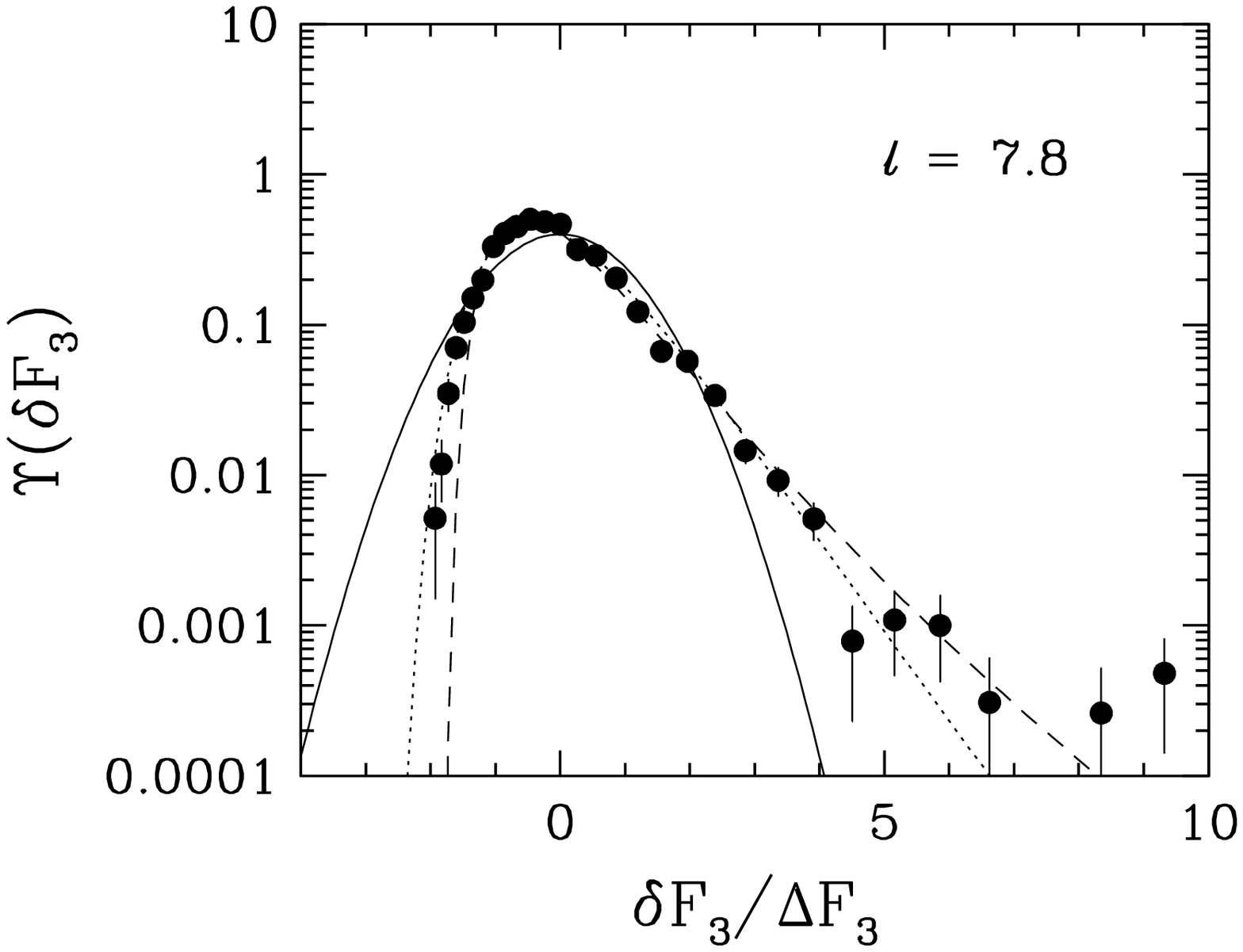,bbllx=35pt,bblly=114pt,bburx=527pt,bbury=492pt,width=4cm}\psfig{figure=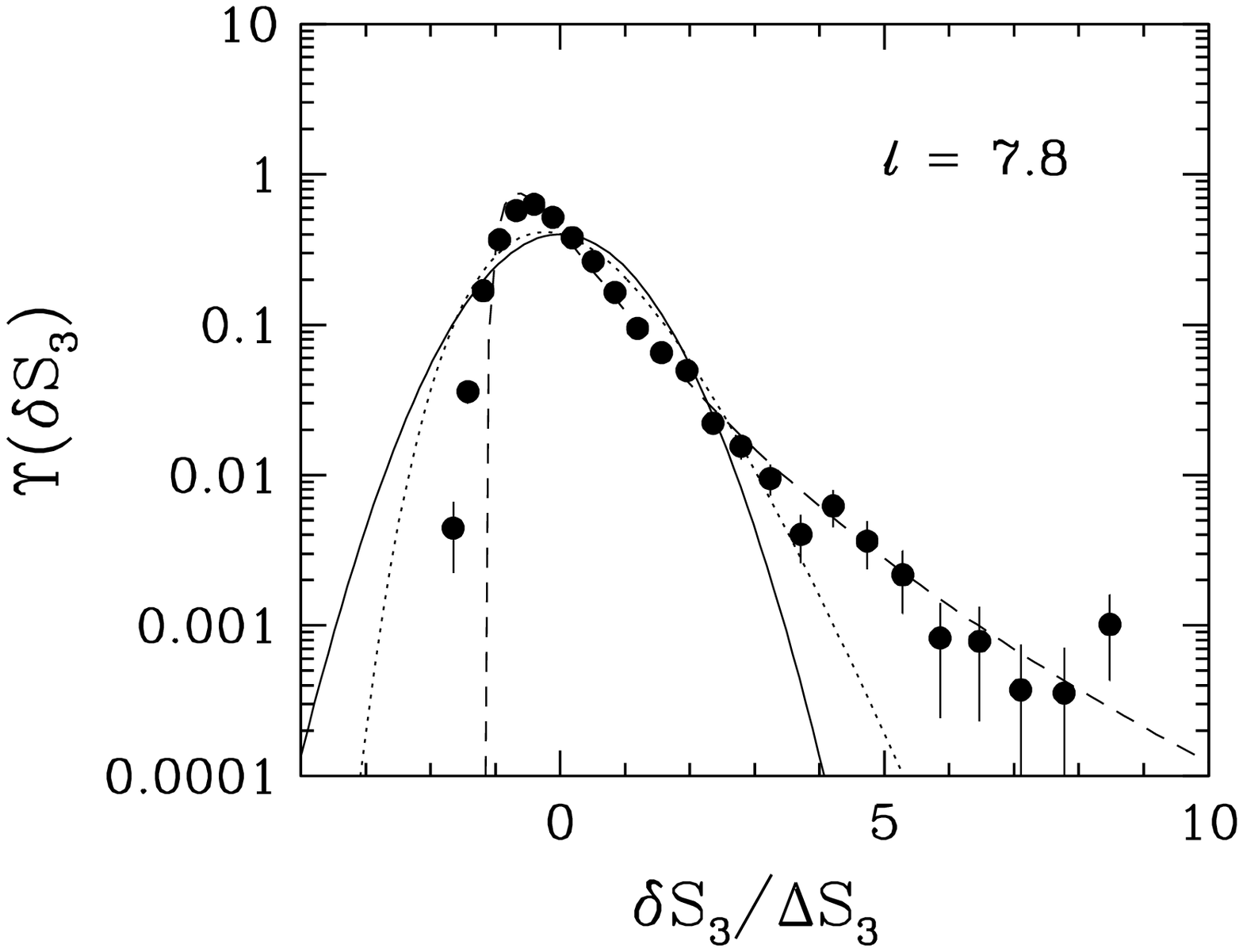,bbllx=35pt,bblly=114pt,bburx=527pt,bbury=492pt,width=4cm}}}
\centerline{\mbox{
\psfig{figure=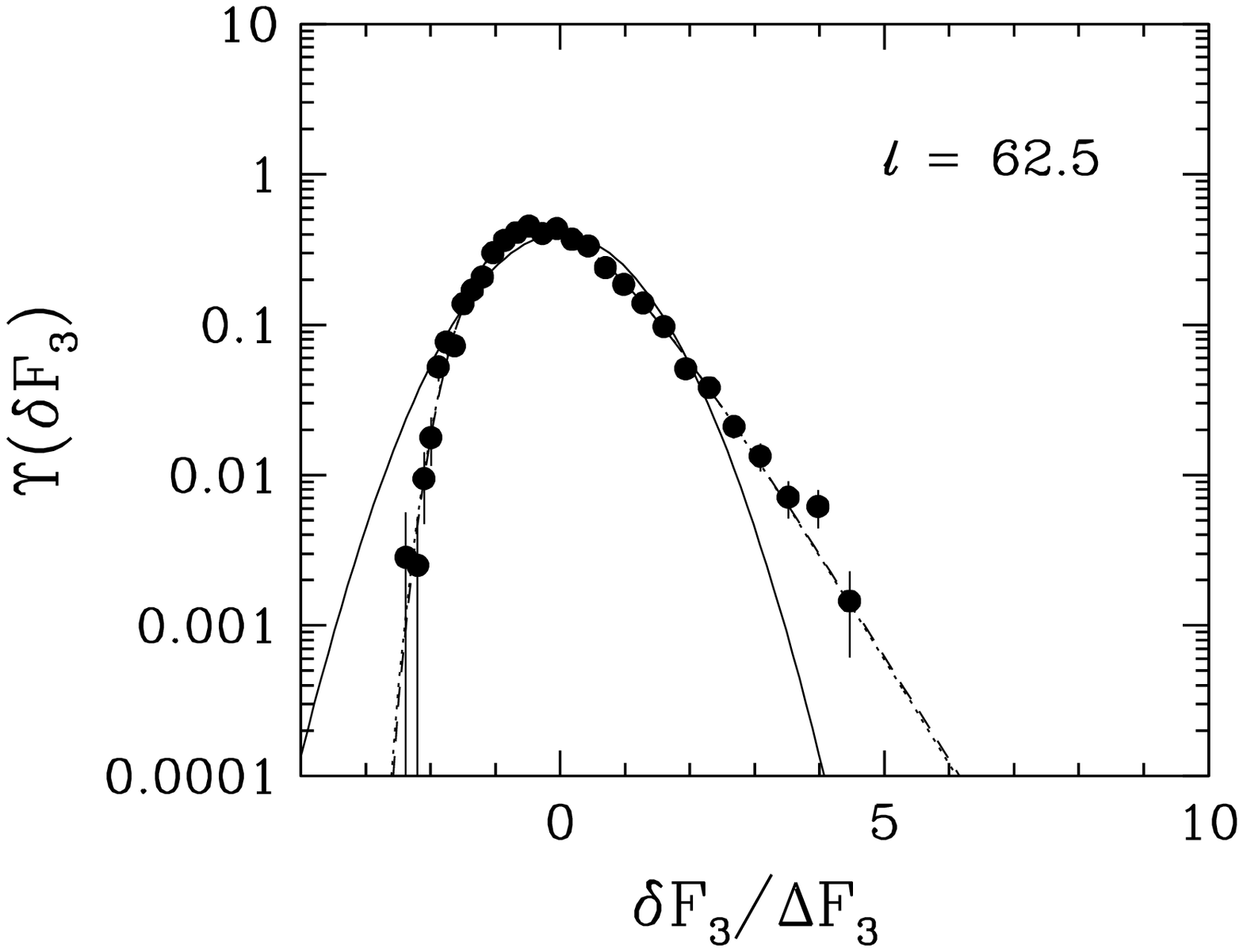,bbllx=35pt,bblly=114pt,bburx=527pt,bbury=492pt,width=4cm}\psfig{figure=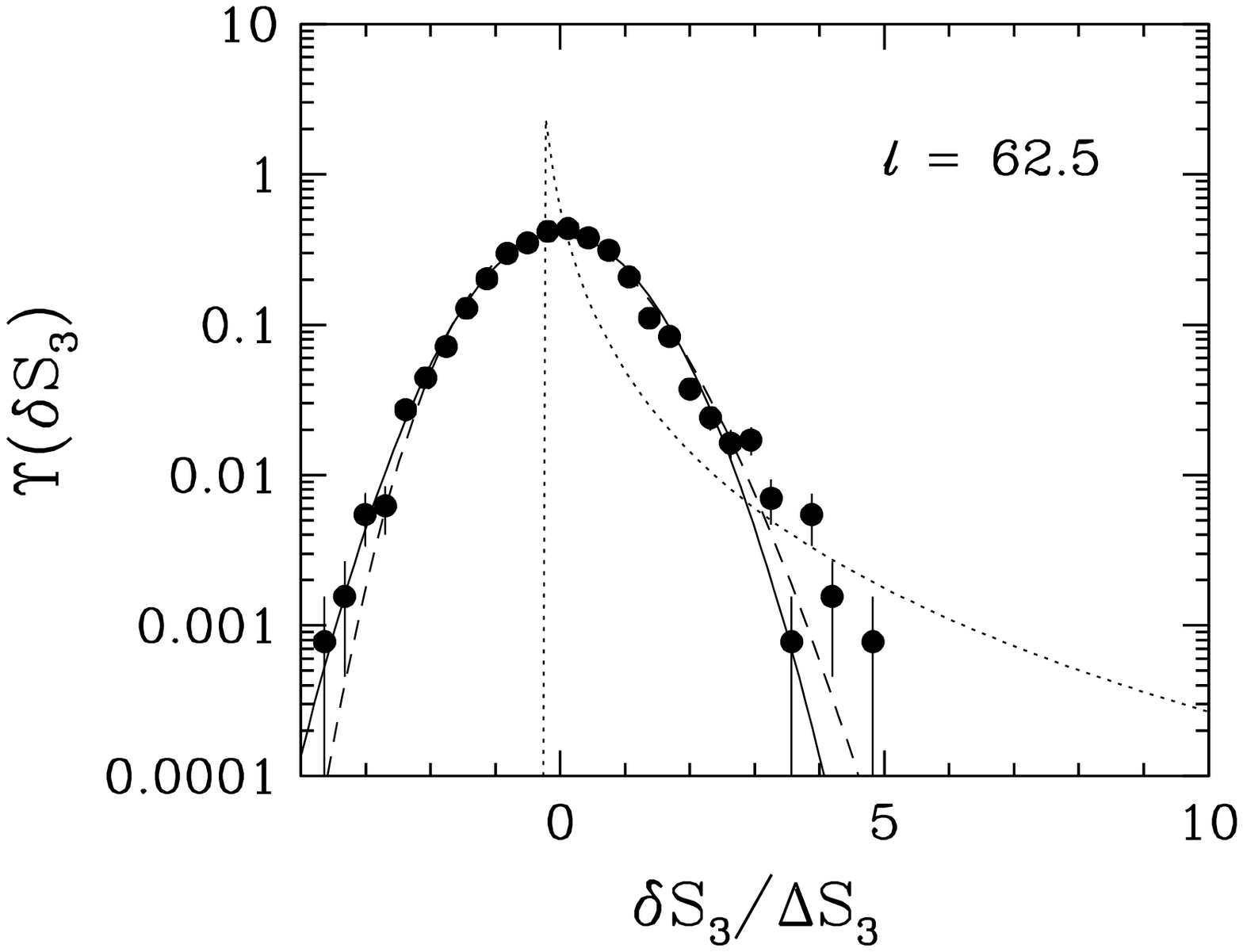,bbllx=35pt,bblly=114pt,bburx=527pt,bbury=492pt,width=4cm}}}
\caption[]{The cosmic distribution function for $F_2$, $\xiav$ (6
upper panels), $F_3$ and $S_3$ (6 lower panels) are shown. Three scales are
considered in each case, indicated in the upper right corner
of each panel, understood in $h^{-1}$ Mpc. The symbols represent
the measurements. The errorbars correspond to the measurement error
due to the fact we use $C_{\cal E}=16^3$ samples as realizations of
our local universe. On each panel, the solid curve, the dots, and the
dashes corresponds to Gaussian, lognormal and extended lognormal as
discussed in the text [equation~(\ref{eq:extln})]. }
\label{fig:figure3}
\end{figure*}

On various scales as indicated on each panel,
Figure~\ref{fig:figure3} shows $\Upsilon(\tA)$ for
$A=F_2$, $\xiav$, $F_3$ and $S_3$ as a function of
the relative difference normalized by the variance, $\delta A/\Delta A$.  
The measurement is compared to the Gaussian limit,
the lognormal distribution (see, e.g., Coles \& Jones 1991), and
a generalized version of it:
\begin{eqnarray}
  \Upsilon(\tA) & = & \frac{s}{\Delta A x  \sqrt{ 2\pi \eta } } \nonumber \\
  & \times & \exp\left[ - \frac{( \ln x-\eta/2 )^2}{2 \eta}
  \right], \label{eq:extln} 
\end{eqnarray} with
$  x =  s (\tA-A)/\Delta A+1$, and 
$ \eta = \ln(1+s^2)$.
The adjustable parameter $s$ is chosen such that the probability
distribution function (\ref{eq:extln}) has same average, variance and skewness
$S = s^3 + 3 s$ than the measured $\Upsilon(\tA)$. Of the three
choices, function (\ref{eq:extln}) was designed to yield
the best fit to the measurements (and it does!) with its three 
adjustable parameters.
Moreover, this approximation also appears to be 
sufficiently accurate to describe the shape of $\Upsilon(\tA)$, 
especially for the large $\tA$ tail. Therefore a third order theory 
is necessary to determine the shape of $\Upsilon(\tA)$
in the studied dynamic range.

The amount of skewness of the dotted curves in
Figure~\ref{fig:figure3} is an indicator
of the magnitude of the cosmic error, since the skewness of the
lognormal distribution is $S = (\Delta A/A)^2 + 3
\Delta A/A$. As we see, function $\Upsilon(\tA)$ is nearly
Gaussian when the cosmic error is small, and in general becomes {\em
increasingly skewed} with $\Delta A/A$. 
For the factorial moments and $\xiav$, the
cosmic distribution function is nearly lognormal, thus for these
estimators the theory of SC, and SBC is sufficient to characterize the
shape of the distribution function.

\section*{Acknowledgments}
SC and IS would like to thank F. Bernardeau, P. Fosalba 
and A. Szalay for stimulating discussions, and 
S. White for warm hospitality
at MPA where a large part of this work was conducted. 
IS was supported by the PPARC rolling grant for 
Extragalactic Astronomy and Cosmology at Durham.
The figures of this paper
were extracted from Colombi, Szapudi \etal (1998).

\end{document}